\documentclass[10pt,letterpaper]{article}
\usepackage{opex3}
\usepackage{graphicx}

\begin{document}

\title{Excitation of surface waves on the interfaces of
general bi-isotropic media}

\author{Seulong Kim and Kihong Kim$^*$}
\address{Department of Energy Systems
Research and Department of Physics, Ajou University, Suwon 16499, Korea}

\email{$^*$khkim@ajou.ac.kr}

\begin{abstract}
We study theoretically the characteristics of surface waves excited at the interface between a metal and a general bi-isotropic medium, which  includes isotropic chiral media and Tellegen media as special cases. We derive an analytical dispersion relation for surface waves, using which we calculate the effective index and the propagation length numerically. We also calculate the absorptance, the cross-polarized reflectance and the spatial distribution of the electromagnetic fields for
plane waves incident on a bilayer system consisting of a metal layer and a bi-isotropic layer in the Kretschmann configuration,
using the invariant imbedding method. The results obtained using the invariant imbedding method agree with those obtained from the dispersion relation perfectly.
In the case of chiral media, the effective index is an increasing function of the chirality index, whereas in Tellegen media,
it is a decreasing function of the Tellegen parameter. The propagation length for surface waves in both cases increase substantially
as either the chirality index or the Tellegen parameter increases. In Tellegen media, it diverges to infinity
when the effective index goes to zero, whereas
in chiral media, it does when the parameters approach the cutoff values where quasi surface waves are excited.
We investigate the characteristics of quasi surface waves excited when the chirality index is sufficiently large.
\end{abstract}

\ocis{(240.6690) Surface waves; (240.6680) Surface plasmons; (160.1585) Chiral media; (160.3918) Metamaterials.}


\section{Introduction}
\label{sec1}

Bi-isotropic media, which include isotropic chiral media and Tellegen media as special cases, are the most general form of linear isotropic media, where the electric displacement {\bf D} and the magnetic induction {\bf B} are related to both the electric field {\bf E} and the magnetic intensity {\bf H} \cite{1}. Using cgs Gaussian units, the constitutive relations for harmonic waves in bi-isotropic media can be written as
\begin{eqnarray}
&&{\bf D}=\epsilon{\bf E}+a{\bf H},\nonumber\\
&&{\bf B}=\mu{\bf H}+a^*{\bf E},
\label{eq:cr}
\end{eqnarray}
where $\epsilon$ is the dielectric permittivity and $\mu$ is the magnetic permeability.
The complex-valued magnetoelectric parameter $a$ is expressed as
\begin{equation}
a=\chi+i\gamma,
\end{equation}
where $\chi$ is the non-reciprocity (or Tellegen) parameter and $\gamma$ is the chirality index.

In uniform bi-isotropic media, right-circularly polarized (RCP) and left-circularly polarized (LCP) waves are the eigenmodes of propagation \cite{1}. In isotropic chiral media, the effective refractive indices for RCP and LCP waves are different, whereas the wave impedance is independent of the helicity \cite{2}. In contrast, in Tellegen media, the effective impedances for RCP and LCP waves are different, whereas the effective refractive index is the same for both waves.

In this paper, we study theoretically surface waves excited on the interface between a metal and a general bi-isotropic medium.
Electromagnetic surface waves in various kinds of media have attracted a large amount of interest among researchers \cite{3}.
Surface plasma waves, or surface plasmons, can be excited on the surface of metals by external
electromagnetic radiation and have been the subject of intense research in recent years due to
their applicability in photonic devices and sensors \cite{4,5}.

Examples of more exotic surface waves in other complex media include Dyakonov waves excited on the interfaces
of anisotropic media \cite {6} and optical Tamm plasmon polaritons excited at the surface of a photonic crystal \cite{7,8}.
Surface polaritons on the surfaces of negative index media have also attracted much attention \cite {9}. Surface waves
due to a spatial inhomogeneity near the surface of inhomogeneous semiconductors
have been studied, too \cite {10}. All of these examples are characterized by the fact that surface waves can be
excited not just by $p$-polarized waves, but by waves of other polarizations.

We aim to extend the study of electromagnetic surface waves to general bi-isotropic media.
We point out that there have been several previous researches devoted to the surface waves
in isotropic chiral media \cite {11,12} and also in topological insulators \cite {13,14}, which, in the presence of weak time-reversal-symmetry-breaking perturbations, can be considered as a kind of Tellegen medium.
At first, we derive a generalized form of the dispersion relation for surface waves at the interface between a metal and a general bi-isotropic medium analytically.
In addition, we use a generalized version of the invariant imbedding method (IIM) for solving wave propagation problems in arbitrarily-inhomogeneous stratified bi-isotropic media, which we have developed recently \cite{15}.
Using this method, we calculate the absorptance, the cross-polarized reflectance and the spatial distribution of the electromagnetic fields for
incident waves of various polarizations in the Kretschmann configuration.
We compare the results obtained using the IIM with those obtained from the analytical dispersion relation and confirm that the agreement is perfect.
In the case of chiral media, the effective index is found to be an increasing function of the chirality index, whereas in Tellegen media,
it is a decreasing function of the Tellegen parameter. The propagation length for surface waves in both cases increase substantially
as either the chirality index or the Tellegen parameter increases. In Tellegen media, it diverges when the effective index goes to zero, whereas
in chiral media, it does when the parameters approach the cutoff values where quasi surface waves are excited.
We investigate the characteristics of quasi surface waves excited when the chirality index is sufficiently large.

In Sec.~\ref{sec2}, we derive the dispersion relation for surface waves in general bi-isotropic media. In Sec.~\ref{sec3}, we give a brief summary of the IIM. In Sec.~\ref{sec4}, we present the results of numerical calculations for surface wave excitation at the interface between a metal and a bi-isotropic medium. Finally, in Sec.~\ref{sec5}, we summarize the paper.

\section{Dispersion relation for surface waves in general bi-isotropic media}
\label{sec2}

\begin{figure}
\centering\includegraphics[width=8cm]{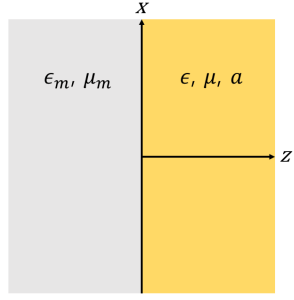}
\caption{Interface between an isotropic medium and a bi-isotropic medium.}
\end{figure}

We consider a plane interface between an ordinary isotropic medium and a bi-isotropic medium located at $z=0$ as illustrated in Fig.~1.
To find a surface wave mode, we look for a solution to Maxwell's equations that propagates in the $x$ direction along the interface but
is exponentially damped away from it.
The isotropic medium has $\epsilon=\epsilon_m$ and $\mu=\mu_m$ and
the bi-isotropic medium is characterized by the parameters $\epsilon$, $\mu$ and $a$ $(=\chi+i\gamma)$. In the $z<0$ region, all field components
are assumed to depend on $x$, $z$ and $t$ as $\exp(\kappa z+iqx-i\omega t)$. In the $z>0$ region corresponding to the bi-isotropic medium, there
are two eigenmodes of circular polarization and the fields can be decomposed as
${\bf E}={\bf E}_+ + {\bf E}_-$ and ${\bf H}={\bf H}_+ + {\bf H}_-$, where ${\bf E}_+$ and ${\bf H}_+$ are proportional to $\exp(-\kappa_+ z+iqx-i\omega t)$, while ${\bf E}_-$ and ${\bf H}_-$ to $\exp(-\kappa_- z+iqx-i\omega t)$.
Using Eq.~(\ref{eq:cr}), Maxwell's curl equations and the condition that $E_x$, $E_y$, $H_x$ and $H_y$ are continuous at $z=0$, we derive
the following dispersion relation for the surface wave mode in a straightforward manner:
\begin{eqnarray}
\left(\frac{\kappa}{\mu_m}+\frac{\kappa_+}{n_+\eta_+ }\right)
\left(\frac{\kappa}{\epsilon_m \eta_-}+\frac{\kappa_-}{n_-}\right)+
\left(\frac{\kappa}{\mu_m}+\frac{\kappa_-}{n_-\eta_-}\right)
\left(\frac{\kappa}{\epsilon_m \eta_+}+\frac{\kappa_+}{n_+}\right)=0,
\label{eq:disp}
\end{eqnarray}
where $n_+$ and $n_-$ ($\eta_+$ and $\eta_-$) are the effective refractive indices (impedances)
for RCP and LCP modes in the bi-isotropic medium, which are given by
\begin{eqnarray}
&&n_+=\left(\epsilon\mu-\chi^2\right)^{1/2}+\gamma,~~~n_-=\left(\epsilon\mu-\chi^2\right)^{1/2}-\gamma,\nonumber\\
&&\eta_+=\left(\frac{\mu}{\epsilon}-\frac{\chi^2}{\epsilon^2}\right)^{1/2}+i\frac{\chi}{\epsilon},
~~~\eta_-=\left(\frac{\mu}{\epsilon}-\frac{\chi^2}{\epsilon^2}\right)^{1/2}-i\frac{\chi}{\epsilon}.
\end{eqnarray}
These formulas are applicable when the real value of $(\epsilon\mu-\chi^2)$ is positive.
The wave vector components $\kappa$, $\kappa_+$ and $\kappa_-$ are defined by
\begin{eqnarray}
\kappa=\left(q^2-k_0^2\epsilon_m\mu_m\right)^{1/2}, ~~~\kappa_+=\left(q^2-k_0^2n_+^2\right)^{1/2},~~~\kappa_-=\left(q^2-k_0^2n_-^2\right)^{1/2},
\label{eq:kap}
\end{eqnarray}
where $k_0$ ($=\omega/c$) is the vacuum wave number. In order to have a pure surface wave mode, all of these quantities have to be positive real numbers.

We assume $\gamma>0$ and $n_+ >n_-$. Then for the range of $q$ such that $ n_- <q/k_0<  n_+$, a coupled mode the LCP component of which is confined
to the interface but the RCP component of which is not is formed. This kind of modes have been termed either quasi surface polariton modes \cite{11} or semi-leaky modes \cite{16}. The condition $q=k_0n_+$ has been called the cutoff condition \cite{12}, which, after some algebraic manipulation, can be written as
\begin{eqnarray}
n_-^2\left(n_+^2-\epsilon_m\mu_m\right)\left(\epsilon\mu-\chi^2\right)^{1/2}=\gamma\left(\epsilon\mu_m+\mu\epsilon_m\right)^2.
\label{eq:cut}
\end{eqnarray}
When any five of the six parameters $\epsilon$, $\mu$, $\chi$, $\gamma$, $\epsilon_m$ and $\mu_m$ are given, this equation can be solved
to give the cutoff value of the other parameter.
In the case where the isotropic medium is a metal, the cutoff value of $\epsilon_m$ can be used to obtain the cutoff frequency $\omega_c$.
Using the Drude expression for the dielectric permittivity, $\omega_c$ can be written as
\begin{eqnarray}
\omega_c=\left(\frac{\omega_p^2}{1-{\rm Re}~\epsilon_m}-\nu^2\right)^{1/2},
\end{eqnarray}
where $\omega_p$ is the plasma frequency and $\nu$ is the collision frequency.

\section{Invariant imbedding method}
\label{sec3}

In our study of surface waves in bi-isotropic media, we use both the analytical dispersion relation derived
in the previous section and the IIM,
which can be applied to more general situations to calculate experimentally relevant quantities.
The IIM for the study of wave propagation in stratified bi-isotropic media has been developed by us recently \cite{15}.
In this section, we give a brief summary of the method.

In stratified media, the parameters $\epsilon$, $\mu$, $\chi$ and $\gamma$
depend on only one spatial coordinate, $z$.
For plane waves propagating in the $xz$ plane, all field components depend on $x$ as $e^{iqx}$.
Starting from Eq.~(\ref{eq:cr}) and Maxwell's curl equations, we derive the
coupled wave equations satisfied by the $y$ components of the fields, $E_y=E_y(z)$ and $H_y=H_y(z)$, in inhomogeneous bi-isotropic media for harmonic waves:
\begin{eqnarray}
{{d^2 \psi}\over{dz^2}}-\frac{d\cal E}{dz}{\cal E}^{-1}(z)
\frac{d\psi}{dz}+\left[k_0^2{\cal E}(z){\cal M}(z)-q^2I\right]\psi=0,
\label{eq:wave}
\end{eqnarray}
where $I$ is a $2\times 2$ unit matrix and
\begin{eqnarray}
\psi=\pmatrix{ E_y \cr H_y \cr},~~{\cal E}=\pmatrix{\mu&-a^*\cr -a&\epsilon\cr},~~
{\cal M}=\pmatrix{\epsilon&a\cr a^*&\mu\cr}.
\label{eq:matrix}
\end{eqnarray}
We assume that the waves are incident from a uniform region ($z>L$)
where $\epsilon=\epsilon_1$ and $\mu=\mu_1$ and
transmitted to another uniform region ($z<0$)
where $\epsilon=\epsilon_2$ and $\mu=\mu_2$.
The inhomogeneous bi-isotropic medium
lies in $0\le z\le L$.
We generalize
Eq.~(\ref{eq:wave}) by replacing the vector wave function
$\psi$ by a $2\times 2$ matrix wave function $\Psi$, the $j$-th
column vector $(\Psi_{1j},\Psi_{2j})^T$ of which represents
the wave function when the incident wave consists only of the $j$-th
wave ($j=1,2$). The index $j=1$ ($j=2$) corresponds to the case where $s$ ($p$) waves are incident.

We are interested in calculating the $2\times 2$ reflection and
transmission coefficient matrices $r=r(L)$ and $t=t(L)$, which we consider as functions of $L$.
In our notation,
$r_{21}$ is the reflection coefficient
when the incident wave is $s$-polarized and the reflected wave is $p$-polarized.
Similarly, $r_{12}$ is the reflection coefficient
when the incident wave is $p$-polarized and the reflected wave is $s$-polarized.
Similar definitions are applied to the transmission coefficients.
The wave functions in the incident and transmitted regions are expressed in
terms of $r$ and $t$:
\begin{eqnarray}
\Psi(z;L)=\left\{ \begin{array}{ll} e^{ip(L-z)}I +e^{ip(z-L)}~r,
&~z>L\\
e^{-ip^\prime z}~t, &~z<0 \end{array} \right., \label{eq:psi}
\end{eqnarray}
where $p$ and $p^\prime$ are the {\it negative} $z$ components of the wave vector
in the incident and transmitted regions.
When $\epsilon_1\mu_1$ and $\epsilon_2\mu_2$ are positive real numbers,
$p$ and $p^\prime$ are obtained from
\begin{eqnarray}
p^2+q^2=k_0^2\epsilon_1\mu_1,~~~{p^{\prime}}^2  +q^2=k_0^2\epsilon_2\mu_2.
\end{eqnarray}
If $\theta$ is the incident angle, $p$, $q$ and $p^\prime$ can be written as
\begin{eqnarray}
p&=&k_0\sqrt{\epsilon_1\mu_1}\cos\theta,~~~
q=k_0\sqrt{\epsilon_1\mu_1}\sin\theta,\nonumber\\
p^\prime&=&\left\{ \begin{array}{ll}
k_0\sqrt{\epsilon_2\mu_2-\epsilon_1\mu_1\sin^2\theta},  &  ~\epsilon_2\mu_2\ge\epsilon_1\mu_1\sin^2\theta \\
ik_0\sqrt{\epsilon_1\mu_1\sin^2\theta-\epsilon_2\mu_2},  &  ~\epsilon_2\mu_2<\epsilon_1\mu_1\sin^2\theta  \end{array} \right..
\label{eq:pqr}
\end{eqnarray}
The invariant imbedding equations satisfied by $r$ and $t$ take the forms
\begin{eqnarray}
&&\frac{dr}{dl}=ip\left(r{\mathcal{E}}+{\mathcal{E}}r\right)
+\frac{ip}{2}\left(r+I\right)\left[{\mathcal{M}}-{\mathcal{E}}+\frac{q^2}{p^2}\left({{\mathcal M}}-{{\mathcal E}}^{-1}\right)\right]\left(r+I\right),\nonumber\\
&&\frac{dt}{dl}=ipt{\mathcal{E}}
+\frac{ip}{2}t\left[{\mathcal{M}}-{\mathcal{E}}+\frac{q^2}{p^2}\left({{\mathcal M}}-{{\mathcal E}}^{-1}\right)\right]\left(r+I\right),
\label{eq:iec}
\end{eqnarray}
where
\begin{eqnarray}
&&{\mathcal E}=\pmatrix{\frac{\mu}{\mu_1}&-\frac{a^*}{\epsilon_1}\cr -\frac{a}{\mu_1}&\frac{\epsilon}{\epsilon_1}\cr}=\pmatrix{\frac{\mu}{\mu_1}&\frac{-\chi+i\gamma}{\epsilon_1}\cr \frac{-\chi-i\gamma}{\mu_1}&\frac{\epsilon}{\epsilon_1}\cr},\nonumber\\
&&{\mathcal M}=\pmatrix{\frac{\epsilon}{\epsilon_1}&\frac{a}{\epsilon_1}\cr \frac{a^*}{\mu_1}&\frac{\mu}{\mu_1}\cr}=\pmatrix{\frac{\epsilon}{\epsilon_1}&\frac{\chi+i\gamma}{\epsilon_1}\cr \frac{\chi-i\gamma}{\mu_1}&\frac{\mu}{\mu_1}\cr}.
\label{eq:matrix2}
\end{eqnarray}
The initial conditions for $r$ and $t$ are
\begin{eqnarray}
&&r_{11}(0)=\frac{\mu_2p-\mu_1 p^\prime}{\mu_2 p+\mu_1 p^\prime},~~r_{22}(0)=\frac{\epsilon_2p-\epsilon_1 p^\prime}{\epsilon_2 p+\epsilon_1 p^\prime},\nonumber\\
&&t_{11}(0)=\frac{2\mu_2p}{\mu_2 p+\mu_1 p^\prime},~~
t_{22}(0)=\frac{2\epsilon_2p}{\epsilon_2 p+\epsilon_1 p^\prime},\nonumber\\
&&r_{12}(0)=r_{21}(0)=t_{12}(0)=t_{21}(0)=0.
\end{eqnarray}
The IIM can also be used in calculating the wave function $\Psi(z;L)$
inside the inhomogeneous medium.
The equation satisfied by $\Psi(z;L)$ takes the form
\begin{eqnarray}
&&\frac{\partial}{\partial l}\Psi(z;l)=ip\Psi(z;L){\mathcal{E}}(l)\nonumber\\&&
~~~~~~~~+\frac{ip}{2}\Psi(z;l)\left\{{\mathcal{M}}(l)-{\mathcal{E}}(l)+\frac{q^2}{p^2}\left[{{\mathcal M}}(l)-{{\mathcal E}}^{-1}(l)\right]\right\}\left[r(l)+I\right].
\label{eq:fe}
\end{eqnarray}
This equation is integrated from $l=z$ to $l=L$ using the initial condition $\Psi(z;z)=I+r(z)$
to obtain $\Psi(z;L)$.

When $p^\prime$ is real and
there is dissipation, the absorptances $A_s$ and $A_p$ for
$s$ and $p$ waves, which are the fractions of the incident wave energy absorbed into the medium, can be written as
\begin{eqnarray}
{A_s}
&=&1-\vert r_{11}\vert^2-\vert \eta_1 r_{21}\vert^2\nonumber\\&&-\frac{\mu_1{\left(\epsilon_2\mu_2-\epsilon_1\mu_1\sin^2\theta\right)}^{1/2}}{\mu_2\sqrt{\epsilon_1\mu_1}\cos\theta}\left(\vert t_{11}\vert^2+\vert \eta_2 t_{21}\vert^2\right),\nonumber\\
{A_p}&=&1-\bigg\vert \frac{1}{\eta_1}r_{12}\bigg\vert^2-\vert r_{22}\vert^2\nonumber\\&&-\frac{\mu_1{\left(\epsilon_2\mu_2-\epsilon_1\mu_1\sin^2\theta\right)}^{1/2}}
{\mu_2\sqrt{\epsilon_1\mu_1}\cos\theta}\left(\bigg\vert \frac{1}{\eta_1}t_{12}\bigg\vert^2+\bigg\vert \frac{\eta_2}{\eta_1}t_{22}\bigg\vert^2\right),
\end{eqnarray}
where $\eta_1=\sqrt{\mu_1/\epsilon_1}$ and $\eta_2=\sqrt{\mu_2/\epsilon_2}$.
If $p^\prime$ is imaginary, ${A_s}$ and ${A_p}$ are given by
\begin{eqnarray}
{A_s}=1-\vert r_{11}\vert^2-\vert \eta_1 r_{21}\vert^2,~~~
{A_p}
=1-\bigg\vert \frac{1}{\eta_1}r_{12}\bigg\vert^2-\vert r_{22}\vert^2.
\end{eqnarray}
In the absence of dissipation, these quantities vanish.

\section{Numerical results}
\label{sec4}

\begin{figure}
\centering\includegraphics[width=9cm]{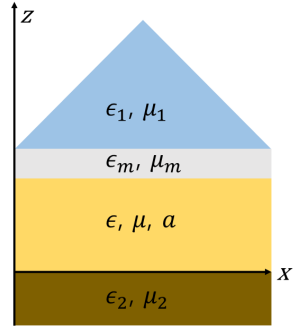}
\caption{Kretschmann configuration used in the invariant imbedding calculations.
A plane wave is incident from the prism with $\epsilon_1$ and $\mu_1$ onto a bilayer system made of a
metal layer with $\epsilon_m$ and $\mu_m$ and a bi-isotropic layer with $\epsilon$, $\mu$ and $a$
and is transmitted to the substrate with $\epsilon_2$ and $\mu_2$ located in $z<0$.}
\end{figure}

\begin{figure}
\centering\includegraphics[width=8cm]{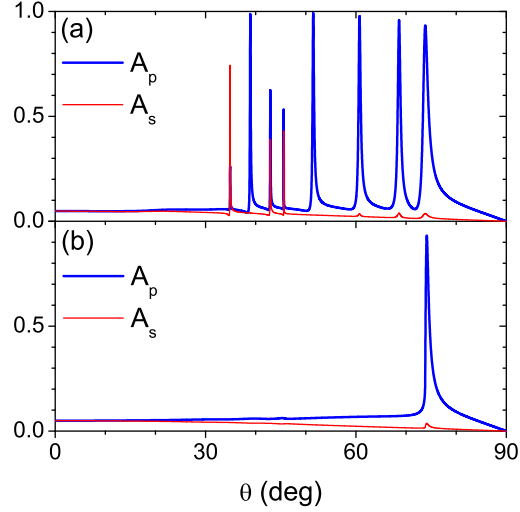}
\caption{Absorptances $A_s$ and $A_p$ for $s$ and $p$ waves of wavelength 633 nm incident on a bilayer system consisting of a silver layer with
$\epsilon_m=-16+i$ and a chiral layer with $\epsilon=2.25$, $\mu=1$ and $\gamma=0.2$ in the Kretschmann configuration plotted versus incident angle. The prism has $\sqrt{\epsilon_1}=1.77$ and the substrate
has (a) $\epsilon_2=\mu_2=1$ and (b) $\epsilon_2=1.5(1.5+\gamma)=2.55$ and $\mu_2=(1.5+\gamma)/1.5\approx 1.13$. The thicknesses of the chiral layer and the silver layer are 1000 nm and 50 nm respectively.}
\end{figure}

\begin{figure}
\centering\includegraphics[width=8cm]{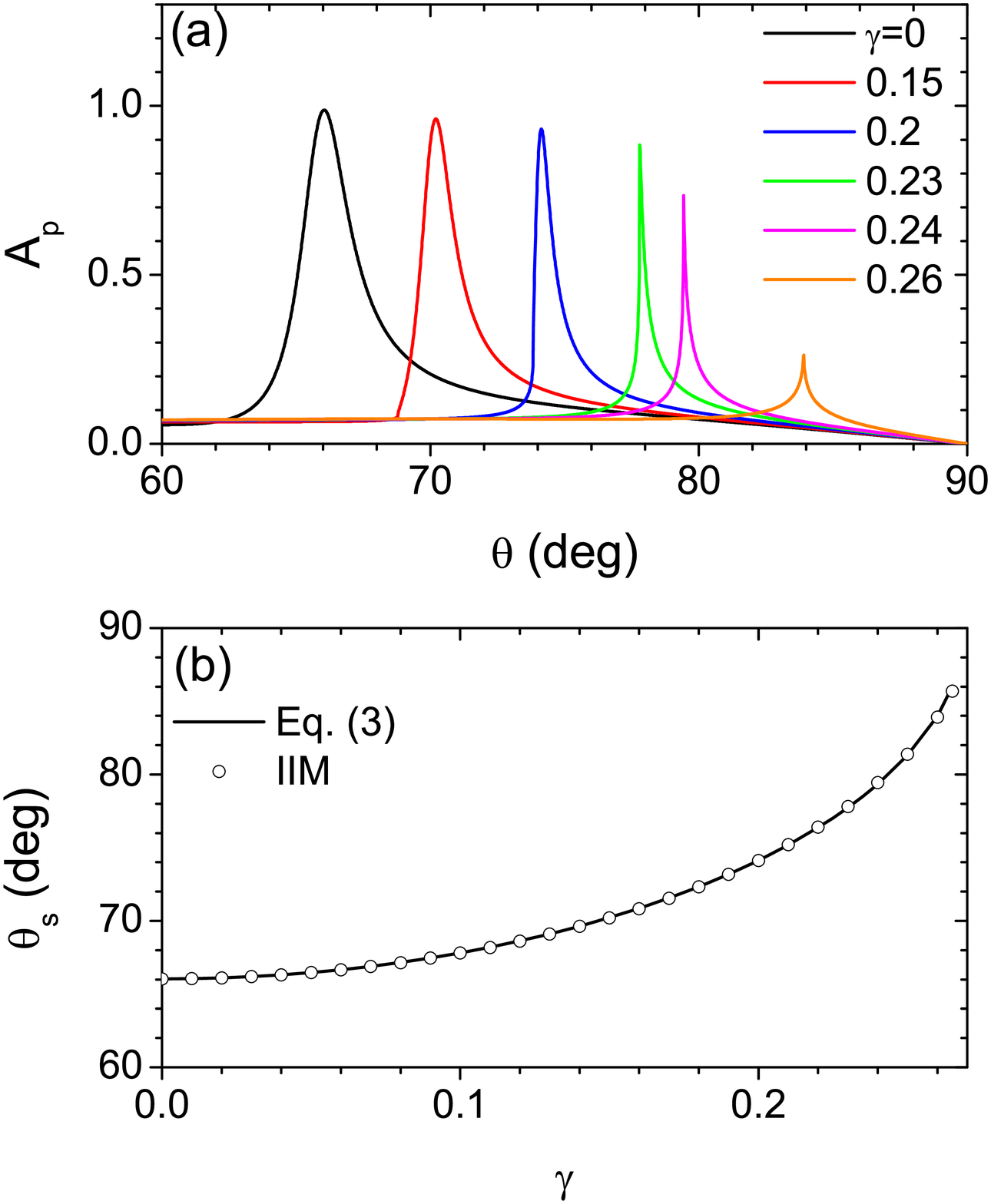}
\caption{(a) Absorptance for $p$ waves of $\lambda=633$ nm incident on a bilayer system consisting of a silver layer with
$\epsilon_m=-16+i$ and a chiral layer with $\epsilon=2.25$, $\mu=1$ and various values of $\gamma$ plotted versus $\theta$. The prism has $\sqrt{\epsilon_1}=1.77$ and the substrate
has $\epsilon_2=1.5(1.5+\gamma)$ and $\mu_2=(1.5+\gamma)/1.5$. The thicknesses of the chiral layer and the silver layer are 1000 nm and 50 nm respectively. (b) Incident angle at which the surface wave is excited, $\theta_s$, for the configuration of the prism/Si layer/chiral layer considered in (a) plotted versus $\gamma$. The result obtained by solving Eq.~(\ref{eq:disp}) is compared with that obtained from the IIM.}
\end{figure}

\begin{figure}
\centering\includegraphics[width=8cm]{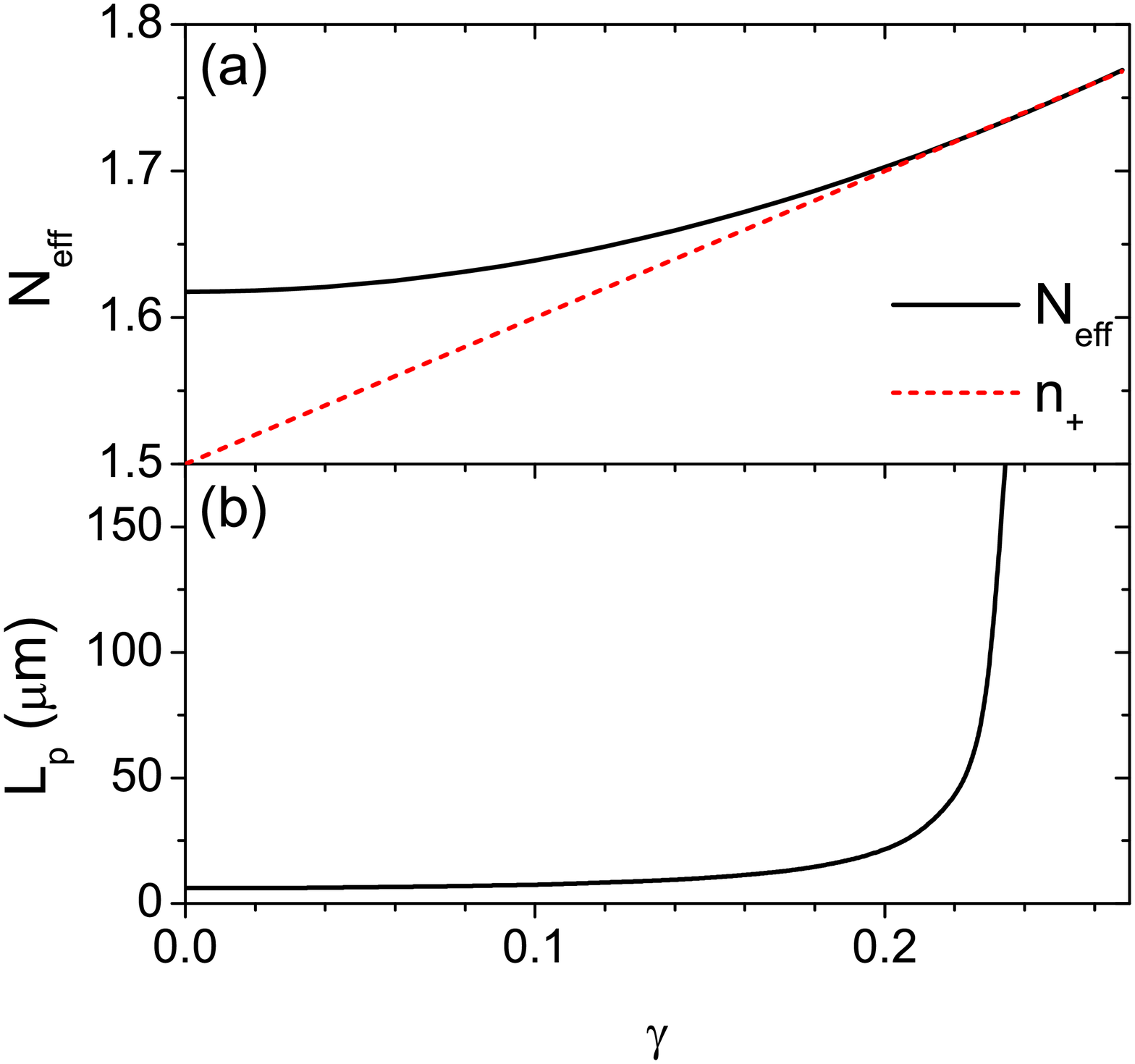}
\caption{(a) Effective index $N_{\rm eff}$ and (b) propagation length $L_p$ for surface waves at the interface
between a metal with $\epsilon_m=-16+i$ and a semi-infinite chiral medium with $\sqrt{\epsilon}=1.5$ and $\mu=1$ plotted versus chirality index $\gamma$. In (a), $N_{\rm eff}$ is compared with $n_+=1.5+\gamma$. As $\gamma$ approaches the cutoff value 0.239, $L_p$ diverges to
infinity.}
\end{figure}

We now present the results of our investigation on the characteristics of surface waves at the interface between a metal and
a bi-isotropic medium using both the IIM and the dispersion relation, Eq.~(\ref{eq:disp}). In the calculations using the IIM, we consider the Kretschmann configuration sketched in Fig.~2, where a plane wave is incident from a prism with $\sqrt{\epsilon_1}=1.77$ and $\mu_1=1$ onto a bilayer system consisting of a silver layer
of thickness $d_m=50$ nm and a bi-isotropic layer of thickness $d_b$ and is transmitted to the substrate located in $z<0$.
We fix the wave frequency such that the vacuum wavelength $\lambda$ is 633 nm, although
the frequency dependence can be studied easily using our method.
At this wavelength, the silver layer
has $\epsilon_m=-16+i$ and $\mu_m=1$. For the bi-isotropic layer, we choose $\epsilon=2.25$, $\mu=1$
and vary the value of $a=\chi+i\gamma$. In order to facilitate a direct comparison with Eq.~(\ref{eq:disp}),
which has been derived for the interface between two semi-infinite media,
the thickness of this layer has to be sufficiently large. In this work, we choose $d_b=1000$ nm.

We first consider the case where the bi-isotropic layer is made of an isotropic chiral medium with $\gamma>0$ and $\chi=0$.
In Fig.~3(a), we plot the absorptances $A_s$ and $A_p$ for $s$ and $p$ waves versus incident angle, when $\gamma=0.2$ and there is no substrate.
On a close examination, we find that both curves display eight peaks at the same angles, even though two of them are too small to be visible in the $s$ wave case. Seven of these eight peaks are associated with the waveguide modes in the thick chiral layer and
only the peak at the largest incident angle $\theta\approx 74.13^\circ$ is due to the excitation of a genuine surface plasmon polariton mode.
In order to see this clearly, we suppress the reflection at the boundary with the substrate by choosing the substrate has $\epsilon_2=1.5(1.5+\gamma)=2.55$ and $\mu_2=(1.5+\gamma)/1.5\approx 1.13$. With this choice, both the chiral layer and the substrate have the same
impedance and the effective index for RCP waves in the chiral layer, $n_+=1.7$, is matched to that of the substrate. The result shown in Fig.~3(b)
demonstrates clearly that this prescription removes all peaks except for the one at $\theta\approx 74.13^\circ$.
That this peak is indeed due to the excitation of a surface wave can be verified directly by solving Eq.~(\ref{eq:disp}), which gives ${\rm Re}~q/k_0\approx 1.7025$. This value corresponds precisely to $\theta\approx 74.13^\circ$.
We also notice that in chiral media, surface waves are excited for both $s$ and $p$ waves unlike in ordinary cases
where only $p$ waves can excite surface plasmons.
In order to concentrate on the surface wave modes, we will use $\epsilon_2=1.5(1.5+\gamma)$ and $\mu_2=(1.5+\gamma)/1.5$ in the rest of calculations.

\begin{figure}
\centering\includegraphics[width=8cm]{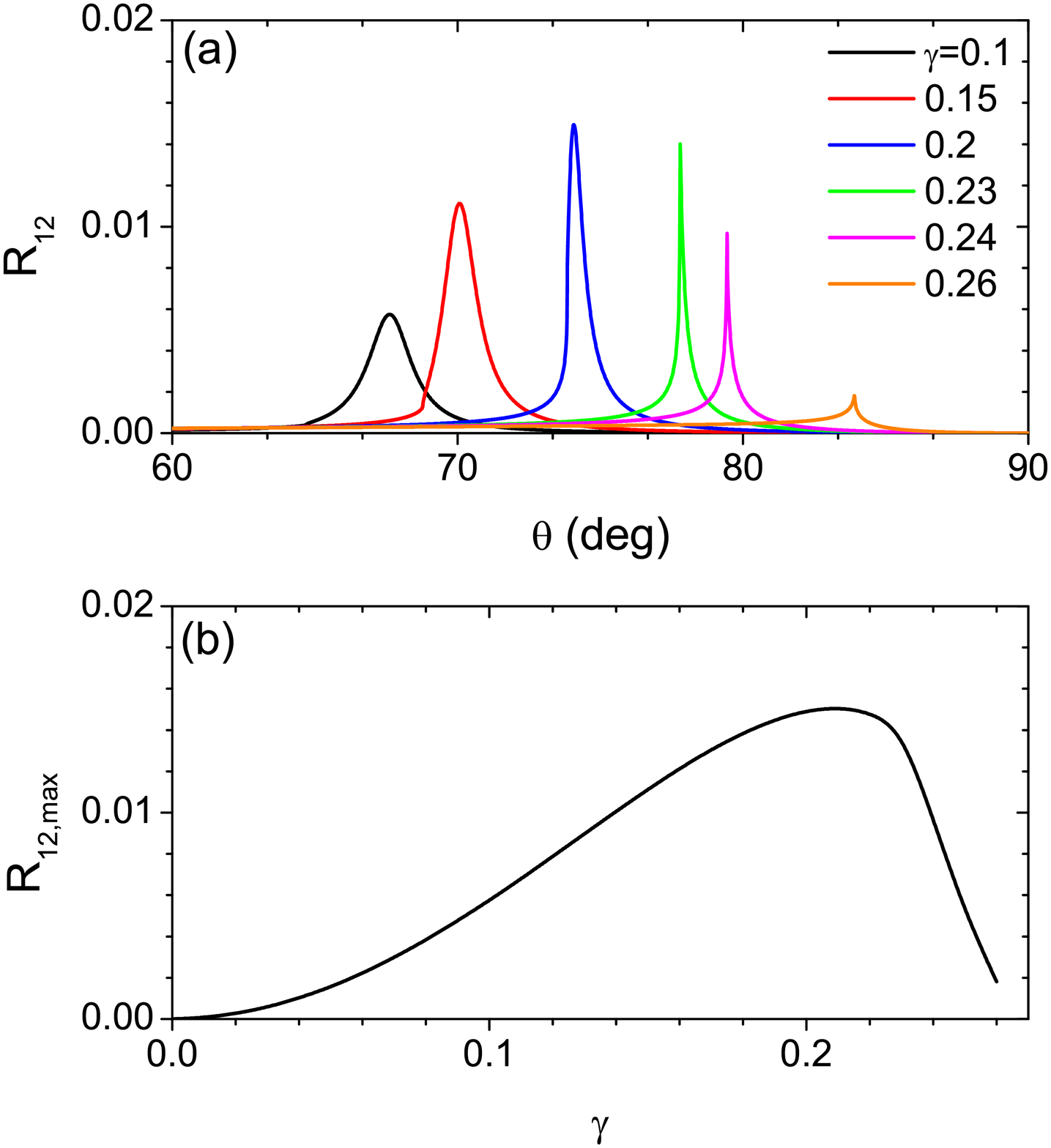}
\caption{(a) Cross-polarized reflectance $R_{12}$ ($=\vert r_{12}\vert^2$) for waves of $\lambda=633$ nm incident on a bilayer system consisting of a silver layer with $\epsilon_m=-16+i$ and a chiral layer with $\epsilon=2.25$, $\mu=1$ and various values of $\gamma$ plotted versus $\theta$. The prism has $\sqrt{\epsilon_1}=1.77$ and the substrate has $\epsilon_2=1.5(1.5+\gamma)$ and $\mu_2=(1.5+\gamma)/1.5$. The thicknesses of the chiral layer and the silver layer are 1000 nm and 50 nm respectively. (b) Maximum value of the cross-polarized reflectance, $R_{12,\rm max}$, versus $\gamma$.}
\end{figure}

\begin{figure}
\centering\includegraphics[width=8cm]{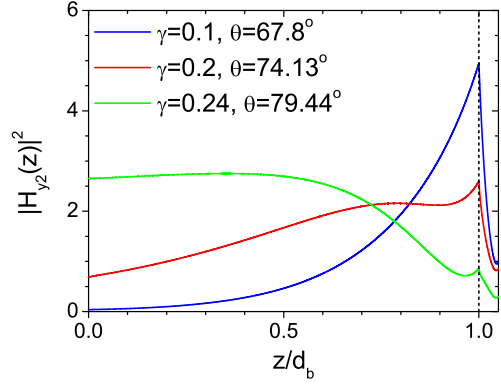}
\caption{Spatial distribution of the intensity of the $y$ component of the magnetic field, $\vert H_{y2}\vert^2$, when $p$ waves of $\lambda=633$ nm are incident on a bilayer system consisting of a silver layer with $\epsilon_m=-16+i$ and a chiral layer with $\epsilon=2.25$, $\mu=1$ and $\gamma=0.1$, 0.2, 0.24. The prism has $\sqrt{\epsilon_1}=1.77$ and the substrate has $\epsilon_2=1.5(1.5+\gamma)$ and $\mu_2=(1.5+\gamma)/1.5$. The thicknesses of the chiral layer and the silver layer are 1000 nm and 50 nm respectively. The incident angle is chosen to be equal to $\theta_s$ for each value of $\gamma$. The dashed vertical line represents the position of the interface.}
\end{figure}

\begin{figure}
\centering\includegraphics[width=8cm]{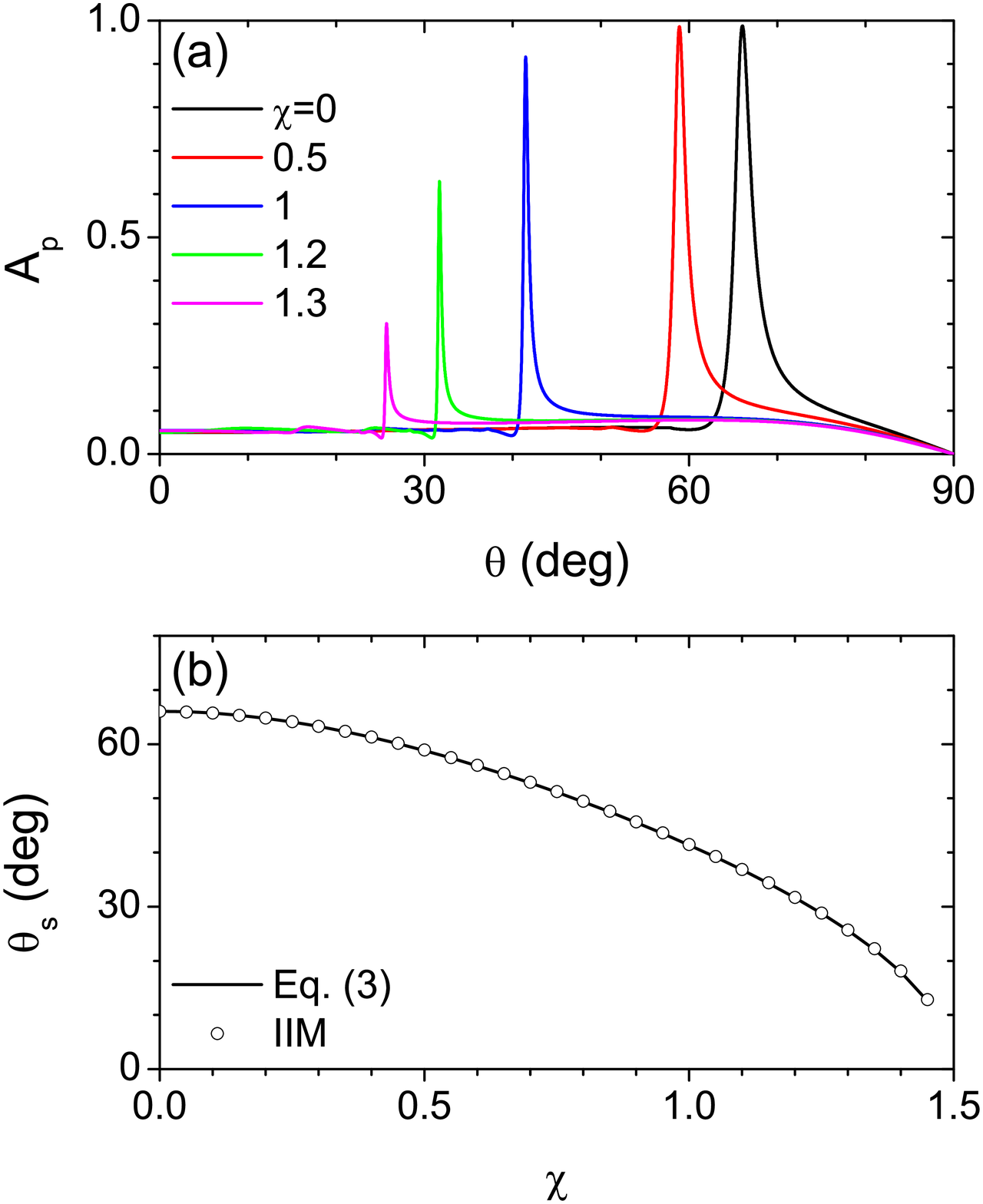}
\caption{(a) Absorptance for $p$ waves of $\lambda=633$ nm incident on a bilayer system consisting of a silver layer with
$\epsilon_m=-16+i$ and a Tellegen layer with $\epsilon=2.25$, $\mu=1$ and various values of $\chi$ plotted versus $\theta$. The prism has $\sqrt{\epsilon_1}=1.77$ and the substrate
has $\epsilon_2=2.25$ and $\mu_2=1$. The thicknesses of the Tellegen layer and the silver layer are 1000 nm and 50 nm respectively. (b) Incident angle at which the surface wave is excited, $\theta_s$, for the configuration of the prism/Si layer/Tellegen layer considered in (a) plotted versus $\chi$. The result obtained by solving Eq.~(\ref{eq:disp}) is compared with that obtained from the IIM.}
\end{figure}

\begin{figure}
\centering\includegraphics[width=8cm]{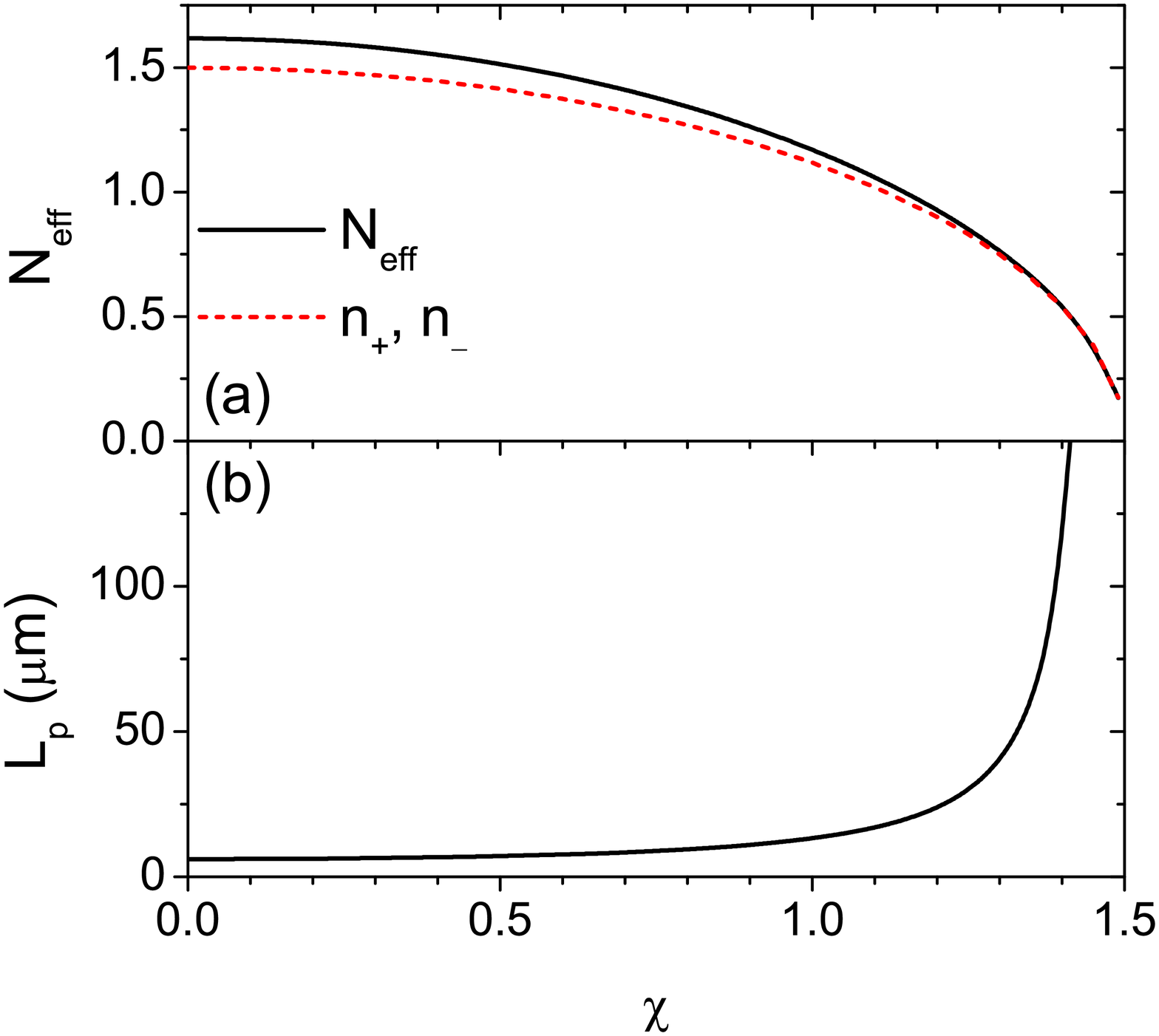}
\caption{(a) Effective index $N_{\rm eff}$ and (b) propagation length $L_p$ for surface waves at the interface
between a metal with $\epsilon_m=-16+i$ and a semi-infinite Tellegen medium with $\epsilon=2.25$ and $\mu=1$ plotted versus Tellegen parameter $\chi$.  In (a), $N_{\rm eff}$ is compared with $n_+=n_-=\left(2.25-\chi^2\right)^{1/2}$. As $\chi$ approaches to 1.5, $N_{\rm eff}$
approaches to zero and $L_p$ diverges to
infinity.}
\end{figure}

\begin{figure}
\centering\includegraphics[width=8cm]{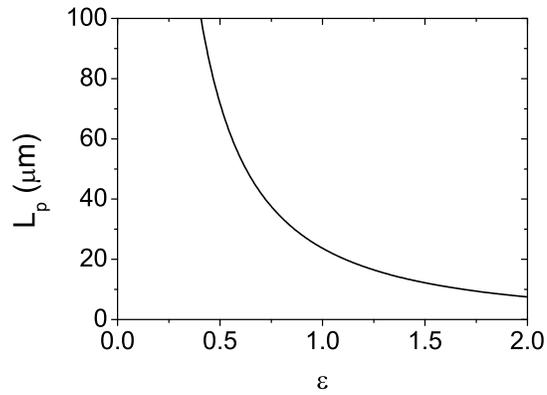}
\caption{Propagation length $L_p$ for surface waves at the interface
between a metal with $\epsilon_m=-16+i$ and a semi-infinite ordinary dielectric medium plotted versus dielectric permittivity $\epsilon$.
As $\epsilon$ approaches to zero, $L_p$ diverges to
infinity.}
\end{figure}

\begin{figure}
\centering\includegraphics[width=8cm]{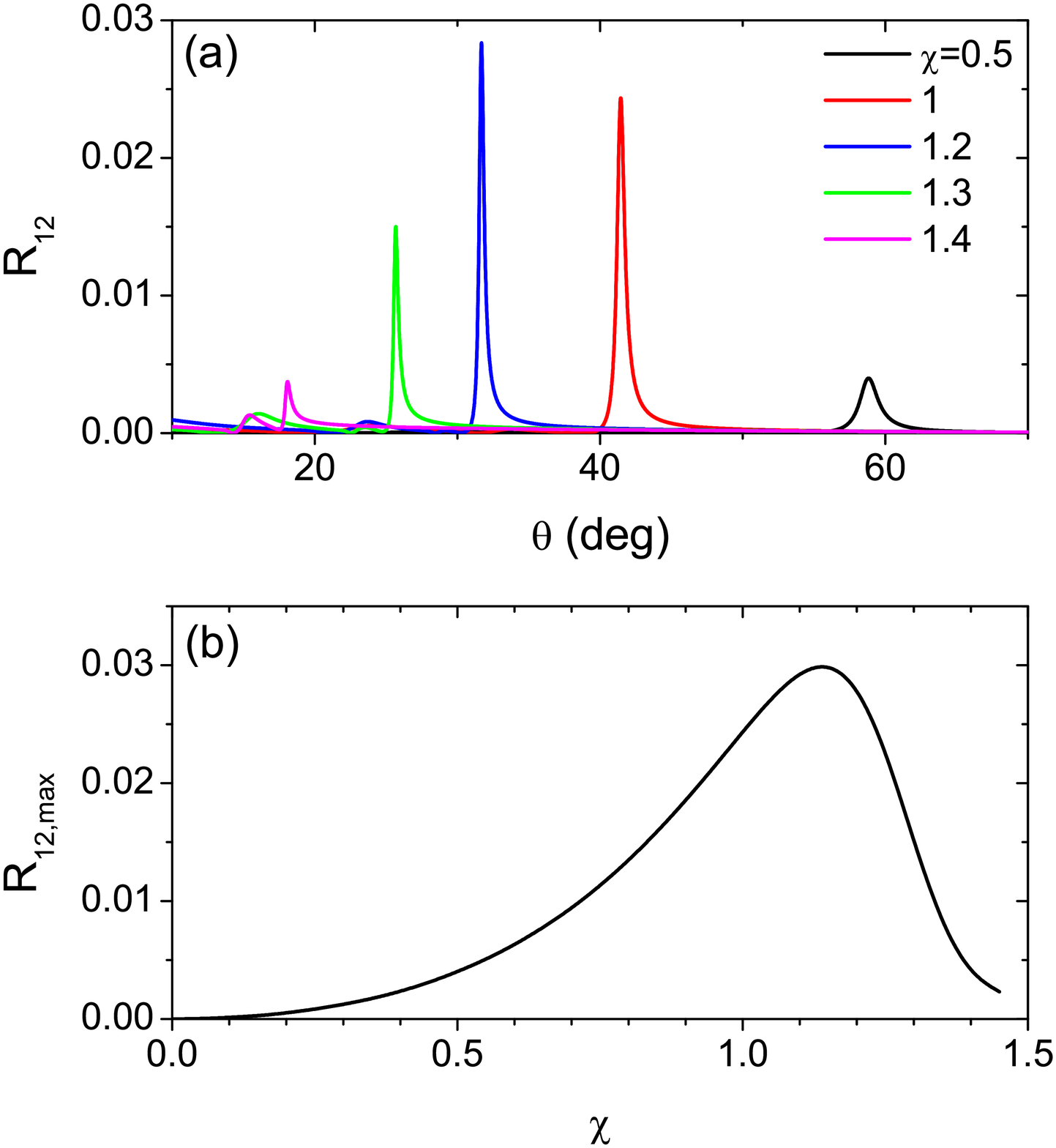}
\caption{(a) Cross-polarized reflectance $R_{12}$ ($=\vert r_{12}\vert^2$) for waves of $\lambda=633$ nm incident on a bilayer system consisting of a silver layer with $\epsilon_m=-16+i$ and a Tellegen layer with $\epsilon=2.25$, $\mu=1$ and various values of $\chi$ plotted versus $\theta$. The prism has $\sqrt{\epsilon_1}=1.77$ and the substrate has $\epsilon_2=1.5$ and $\mu_2=1$. The thicknesses of the chiral layer and the silver layer are 1000 nm and 50 nm respectively. (b) Maximum value of the cross-polarized reflectance, $R_{12,\rm max}$, versus $\chi$.}
\end{figure}

\begin{figure}
\centering\includegraphics[width=8cm]{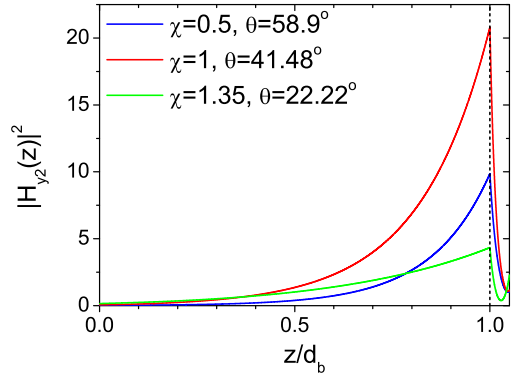}
\caption{Spatial distribution of the intensity of the $y$ component of the magnetic field, $\vert H_{y2}\vert^2$, when $p$ waves of $\lambda=633$ nm are incident on a bilayer system consisting of a silver layer with $\epsilon_m=-16+i$ and a Tellegen layer with $\epsilon=2.25$, $\mu=1$ and $\chi=0.5$, 1, 1.35. The prism has $\sqrt{\epsilon_1}=1.77$ and the substrate has $\epsilon_2=2.25$ and $\mu_2=1$. The thicknesses of the chiral layer and the silver layer are 1000 nm and 50 nm respectively. The incident angle is chosen to be equal to $\theta_s$ for each value of $\chi$. The dashed vertical line represents the position of the interface.}
\end{figure}

\begin{figure}
\centering\includegraphics[width=8cm]{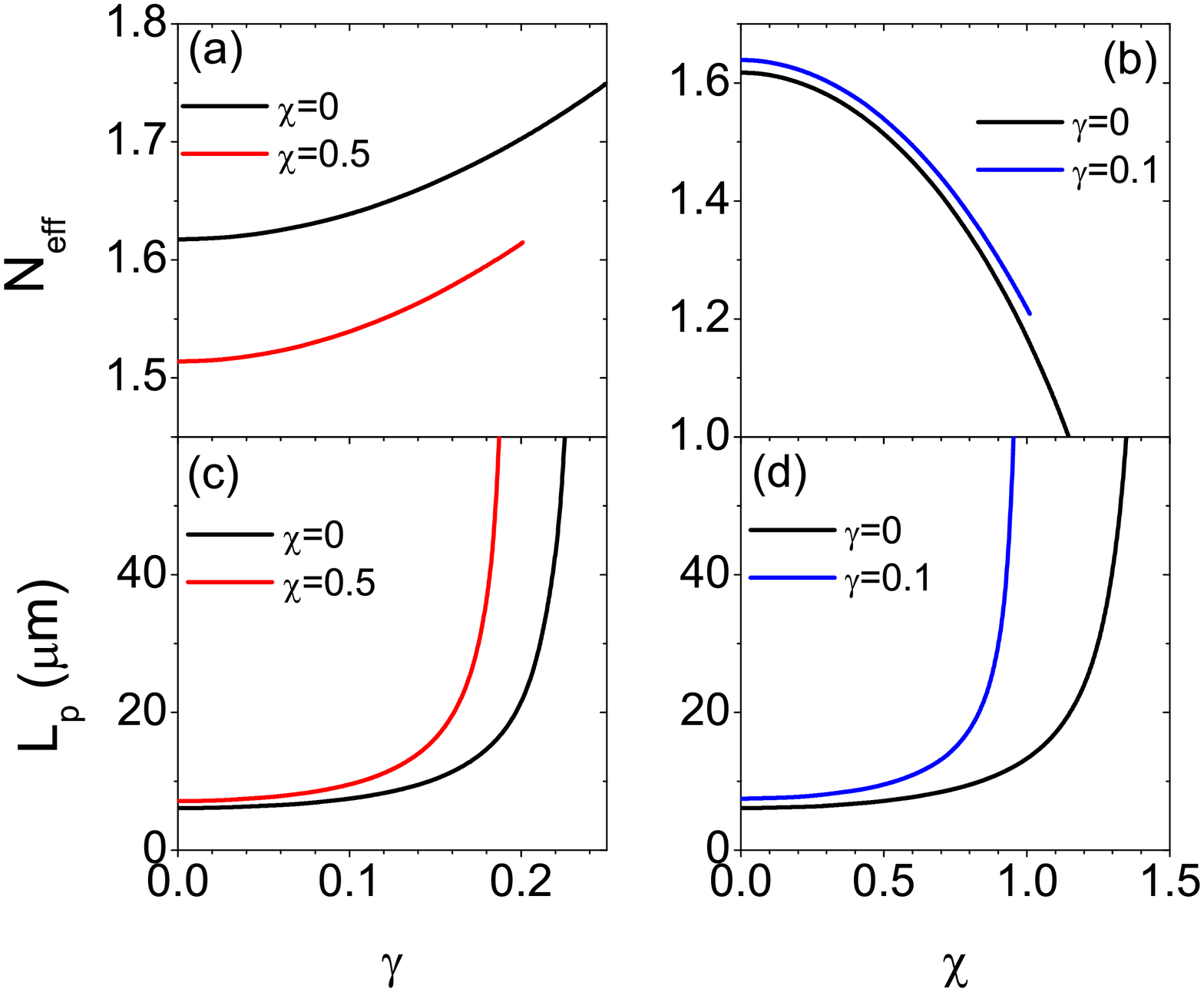}
\caption{(a) Effective index and (c) propagation length for surface waves at the interface
between a metal with $\epsilon_m=-16+i$ and a semi-infinite bi-isotropic medium with $\epsilon=2.25$, $\mu=1$ and $\chi=0.5$ plotted versus chirality index $\gamma$ and compared with the results obtained when $\chi=0$. The cutoff value for $\gamma$ is approximately 0.2, where
$N_{\rm eff}$ diverges to infinity. (b) Effective index and (d) propagation length for surface waves at the interface
between a metal with $\epsilon_m=-16+i$ and a semi-infinite bi-isotropic medium with $\epsilon=2.25$, $\mu=1$ and $\gamma=0.1$ plotted versus Tellegen parameter $\chi$ and compared with the results obtained when $\gamma=0$. The cutoff value for $\chi$ is approximately 1.025, where
$N_{\rm eff}$ diverges to infinity. }
\end{figure}

In Fig.~4(a), we show the $p$ wave absorptance for several different values of $\gamma$. We find that as $\gamma$ increases, the incident angle at which the surface wave is excited, $\theta_s$, shifts to larger angles and the maximum value of $A_p$ decreases. In Fig.~4(b),
we compare the values of $\theta_s$ obtained using the IIM with those obtained using the analytical dispersion
relation. The agreement is seen to be excellent. The effective refractive indices for the chiral layer in the present case are $n_\pm=1.5\pm \gamma$,
therefore surface waves cannot be excited for $\gamma\ge 0.27$ when the wave is incident from a prism with $\sqrt{\epsilon}=1.77$. By solving Eq.~(\ref{eq:cut}) numerically, we find that the cutoff value for $\gamma$ is approximately 0.239. When $\gamma<0.239$, both RCP and LCP components are
confined to the interface and we have a true surface plasmon polariton mode. For $0.239<\gamma<0.27$, however, a quasi surface mode the LCP component of which is confined
to the interface but the RCP component of which is not is formed. When we solve Eq.~(\ref{eq:disp}) in this parameter region,
we have to use $\kappa_+=i(k_0^2n_+^2-q^2)^{1/2}$ instead of the expression given in Eq.~(\ref{eq:kap}). We find that even for quasi surface modes, the result obtained using the IIM shows a sharp
absorption peak at the same value of $\theta_s$ calculated using Eq.~(\ref{eq:disp}).

By a numerical solution of Eq.~(\ref{eq:disp}), we obtain the real and imaginary parts of the wave vector component $q$. We define the effective index $N_{\rm eff}$ by $N_{\rm eff}={\rm Re}~q/k_0$ and the propagation length $L_p$ by $L_p = 1/(2{\rm Im}~q)$ \cite{12}.
As shown in Fig.~5, $N_{\rm eff}$ is an increasing function of $\gamma$, which is equivalent to saying that $\theta_s$ is an increasing function of $\gamma$. The propagation length also increases monotonically as $\gamma$ increases.
As $\gamma$ approaches the cutoff value 0.239, $N_{\rm eff}$ approaches $n_+$ and $L_p$ diverges to
infinity, which has been pointed out previously in \cite{12}.

Using the IIM, we have also calculated the cross-polarized reflectance $R_{12}$ ($=\vert r_{12}\vert^2$), which is a measure of
the conversion from $p$ to $s$ waves. In Fig.~6(a), we plot $R_{12}$ obtained for various values of $\gamma$ versus $\theta$.
We find that $R_{12}$ shows a sharp peak at an angle very close to $\theta_s$. In Fig.~6(b), we plot
the maximum value of the cross-polarized reflectance, $R_{12,\rm max}$, as a function of $\gamma$.
As $\gamma$ increases, it increases initially and then decreases toward zero
as $\gamma$ approaches the cutoff value. We have verified that in the region where $\gamma$ is sufficiently small, $R_{12,\rm max}$
is proportional to $\gamma^2$.

In Fig.~7, we show the spatial distribution of the intensity of the $y$ component of the magnetic field, $\vert H_{y2}(z)\vert^2$, when $p$ waves are
incident on the the bilayer system. The incident angle is chosen to be equal to $\theta_s$ for each value of $\gamma$. For small values of $\gamma$, we find that a true surface wave mode is excited at the interface. As $\gamma$ increases, the field penetrates more and more deeply into the chiral layer. When $\gamma$ is above the cutoff value, we
observe that the field has a weak component confined to the interface and a propagating component which is not confined to it.

Next we shift to the case where the bi-isotropic medium is a non-reciprocal Tellegen medium with $\chi>0$ and $\gamma=0$.
In this case, we have $n_+=n_-=\left(\epsilon\mu-\chi^2\right)^2$ and $\eta_+\ne\eta_-$. Since $\kappa_+=\kappa_-$,
there is no cutoff value for $\chi$ and quasi surface polariton modes do not exist.
In the calculation using the IIM, we consider a bilayer system with the substrate having $\epsilon_2=2.25$ and $\mu_2=1$
to remove all waveguide modes.
In Fig.~8(a), we show the $p$ wave absorptance for several different values of $\chi$ obtained using the IIM. In contrast to
the case with a chiral layer, we find that as $\chi$ increases, the incident angle at which the surface wave is excited, $\theta_s$, shifts to smaller angles. In Fig.~8(b),
we compare the values of $\theta_s$ obtained using the IIM with those obtained using Eq.~(\ref{eq:disp}). The agreement is extremely good. The effective refractive indices for the Tellegen layer in the present case are $n_+=n_-=\left(2.25-
\chi^2 \right)^{1/2}$,
therefore surface waves are expected to arise when $\chi<1.5 $.

In Fig.~9, we plot the effective index $N_{\rm eff}$ and the propagation length $L_p$ obtained by solving Eq.~(\ref{eq:disp}) numerically versus $\chi$.
In contrast to the chiral case, $N_{\rm eff}$ is a decreasing function of $\chi$, which is equivalent to saying that $\theta_s$ is a decreasing function of $\chi$. Similarly to the chiral case, however, the propagation length increases monotonically as $\chi$ increases.
As $\chi$ approaches $\sqrt{\epsilon\mu}=1.5$, $N_{\rm eff}$ approaches $n_+=n_-=\left(2.25-\chi^2\right)^{1/2}\approx 0$ and $L_p$ diverges to
infinity. The divergence of $L_p$ in the present case is due to the vanishing of the effective index $N_{\rm eff}$. As will be shown in Fig.~12,
the electromagnetic fields decay very slowly in the dielectric side of the metal-dielectric interface, when the effective index of the dielectric is close to zero. This makes the surface waves propagate mostly along the dielectric and reduces the propagation loss greatly. A similar phenomenon occurs in the surface plasmons between a metal and an ordinary dielectric medium with very small positive dielectric permittivity. In Fig.~10,
we show the propagation length $L_p$ for surface waves at the interface
between a metal with $\epsilon_m=-16+i$ and a semi-infinite ordinary dielectric medium as a function of the dielectric permittivity $\epsilon$.
As $\epsilon$ approaches to zero, $L_p$ is shown to diverge.

Using the IIM, we have also calculated the cross-polarized reflectance $R_{12}$. In Fig.~11(a), we plot $R_{12}$ obtained for various values of $\chi$ versus incident angle.
We find that $R_{12}$ shows a sharp peak at an angle very close to $\theta_s$. In Fig.~11(b), we plot
the maximum value of the cross-polarized reflectance, $R_{12,\rm max}$, as a function of $\chi$.
The qualitative behavior is rather similar to that of the chiral case.
As $\chi$ increases, $R_{12,\rm max}$ increases initially and then decreases toward zero
as $\chi$ approaches 1.5. Analogously to the chiral case, we have verified that in the region where $\chi$ is sufficiently small, $R_{12,\rm max}$
is proportional to $\chi^2$.

In Fig.~12,
the spatial distribution of the intensity of the $y$ component of the magnetic field, $\vert H_{y2}(z)\vert^2$, when $p$ waves are
incident on the the bilayer system. The incident angle is chosen to be equal to $\theta_s$ for each value of $\chi$. For all cases, we find that a true surface wave mode is excited at the interface. As $\chi$ increases, the field penetrates more deeply into the Tellegen layer.

Finally, we consider the case of general bi-isotropic media with nonzero $\chi$ and $\gamma$. In Figs.~13(a) and 13(c), the effective index and the propagation length for surface waves at the interface
between a metal with $\epsilon_m=-16+i$ and a semi-infinite bi-isotropic medium with $\epsilon=2.25$, $\mu=1$ and $\chi=0.5$ are plotted versus chirality index $\gamma$ and compared with the results obtained when $\chi=0$. The cutoff value for $\gamma$ is approximately 0.2, where
$N_{\rm eff}$ diverges to infinity. In Figs.~13(b) and 13(d), the effective index and the propagation length for surface waves at the interface
between a metal with $\epsilon_m=-16+i$ and a semi-infinite bi-isotropic medium with $\epsilon=2.25$, $\mu=1$ and $\gamma=0.1$ are plotted versus Tellegen parameter $\chi$ and compared with the results obtained when $\gamma=0$. The cutoff value for $\chi$ is approximately 1.025, where
$N_{\rm eff}$ diverges to infinity. The effective index is an increasing function of $\gamma$ in Fig.~13(a), whereas it is
a decreasing function of $\chi$ in Fig.~13(b). We note that in both cases, the propagation length
is substantially enhanced when both $\gamma$ and $\chi$ are nonzero.

\section{Conclusion}
\label{sec5}

In this paper, we have studied theoretically the characteristics of the surface waves excited at the interface between a metal and a general bi-isotropic medium. We have derived an analytical dispersion relation for surface waves, using which we have calculated the effective index and the propagation length numerically. In addition, we have calculated the absorptance, the cross-polarized reflectance and the spatial distribution of the electromagnetic fields for
incident waves of various polarizations in the Kretschmann configuration, using a generalized version of the IIM.
We have compared the results obtained using the IIM with those obtained from the analytical dispersion relation and found that the agreement is perfect.
We have found that in chiral media, the effective index is an increasing function of the chirality index, whereas in Tellegen media,
it is a decreasing function of the Tellegen parameter. The propagation length for surface waves in both cases has been found to increase substantially
as the bi-isotropic parameter increases. In Tellegen media, it diverges when the effective index goes to zero, whereas
in chiral media, it does when the parameters approach the cutoff values where quasi surface waves are excited.

It is straightforward to apply our method to more general bi-isotropic media with nonzero $\gamma$ and $\chi$.
It is also possible to generalize our method to surface waves at the interface between two different kinds of bi-isotropic media including those with
negative refractive indices. A deep investigation of these and related problems may lead to the development of useful polarization-sensitive photonic devices.

\section*{Acknowledgments}
This work has been supported by the National Research Foundation of Korea Grant (NRF-2015R1A2A2A01003494) funded by the Korean Government.

\end{document}